\documentclass[%
reprint,
superscriptaddress,
nofootinbib,
nobibnotes,
noeprint,
amsmath,amssymb,
aps,
prl,
longbibliography
]{revtex4-2}

\usepackage{graphicx}
\usepackage{dcolumn}
\usepackage{siunitx}
\usepackage{bm}
\usepackage{upgreek}

\usepackage{graphicx}
\usepackage{float}
\usepackage{caption}
\usepackage{subcaption}
\usepackage{orcidlink}
\usepackage{xcolor}
\usepackage{color,soul} 
\usepackage{braket}

\graphicspath{{figures/}}

\begin{document}
\preprint{APS/123-QED}

\title{A large-momentum-transfer Raman atom interferometer without $k$-reversal}

\author{Guanchen Peng\orcidlink{0009-0009-9583-3489
}}
\author{Bryony Lanigan\orcidlink{0009-0009-7292-4185}}
\author{R. Shah\orcidlink{0000-0002-7884-1534}}
\author{J. Lim\orcidlink{0000-0002-1803-4642}}
\author{A. Kaushik\orcidlink{0000-0002-6046-9930}}
\author{J. P. Cotter\orcidlink{0000-0002-7055-0206}}
\author{E. A. Hinds\orcidlink{0000-0002-0824-9834}}
\author{B. E. Sauer\orcidlink{0000-0002-3286-4853}}

\affiliation{Centre for Cold Matter, Blackett Laboratory, Imperial College London,  Prince Consort Road, London SW7 2AZ, United Kingdom.}
\date{\today}

\begin{abstract}
We present a Raman  atom interferometer using large momentum transfer without reversing the direction of the effective wavevector ($k$-reversal). More specifically, we use a microwave $\pi$/2 pulse to manipulate the spin state of $^{87}$Rb atoms before applying a Raman light $\pi$ pulse to achieve 4$\hbar k$  momentum transfer per Raman light pulse. A microwave $\pi$ pulse in the middle of the interferometer sequence reverses the spin states, which allows closing of the interferometer arms by the same Raman light $\pi$ pulses without propagation reversal. We present a proof-of-principle demonstration of a 4$\hbar k$ large-momentum-transfer (LMT) atom interferometer and discuss its scalability. Our results extend the scope of using LMT atom optics. 
\end{abstract}

\maketitle{}


Atom interferometers have emerged as pivotal tools in modern physics, enabling high-precision measurements of fundamental constants \cite{Kasevich_bigG_measurement_2007, Rosi2014_bigGmeasurment, Bouchendira_FineStructmeasurement_2011, HMuller_fineStructmeasure_2018, Morel2020_FoneStructureConst}, rigorous tests of physical laws \cite{Rasel_QuantumFreeFall_2014, Zhou_EquivTestDoubleDiff_2015, Zhang2023_LorentzViolationTest, KasevichGroup2022_ABeffect}, and ambitious efforts to detect dark energy~\cite{burrage2015,panda2024}, dark matter or gravitational waves \cite{Canuel2018_MIGA, Zhan2020_ZAIGA, badurina2020aion,abe2021matter}.

In these measurements, light is used to split an atomic wave packet, which then propagates along two distinct interferometer arms, before being recombined to obtain interference fringes that are sensitive to the acceleration of the atom. The exchange of momentum with the light can entangle the two  momentum states of the atom with two  different internal states, simplifying detection of the fringes  by eliminating the need to resolve the two output ports spatially. Raman transitions between hyperfine ground states have proved to be an effective way of doing this  in technological applications including inertial sensing \cite{KasevichGroup2013_InertialLaunching, d’ArmagnacdeCastanet2024_rotationMeasurements, geiger2011_InertialNavigation, cheiney2018_InertialNavigation, Templier2022_InertialSensing} and gravity mapping \cite{Aisha2022_gravityGradiometer, Li2023_portableGravimeter, menoret2018_gravimeter, wu2019gravimeter}. 

The acceleration sensitivity of an atom interferometer increases with the space-time area enclosed by the arms. One method to increase sensitivity, therefore, is to increase the propagation time. In free-fall (as opposed to trapped geometries  e.g.\cite{panda2024coherence}), propagation times have been extended by the use of atomic fountains \cite{Kasevich1992_atomicFountain, KasevichGroup2013_InertialLaunching, clairon1991ramsey_fountain}, drop towers \cite{kulas2017miniaturized_BremenDropTower, Mutinga2013_dropTower} and other micro-g environments \cite{Williams2024_spaceAI, Lachmann2021_space, elsen2023_MAIUS, barrett2016_inAPlane}. 
These approaches allow longer free-fall times, but at the expense of large infrastructure, increasing the scale and complexity of experiments. An alternative is to increase the  momentum imparted to the atom by the beam splitter. Large-momentum-transfer (LMT) methods have achieved differential momenta as high as $16 \hbar k$ using spin-dependent Raman kicks \cite{HMuller_SDK_LMT_2018}, $100\hbar k$ using sequential Bragg atom optics \cite{Kasevich_102hk_2011}, and $ 400\hbar k$ using a combination of Bloch oscillations and Bragg transitions in twin optical lattices \cite{gebbe2021twin}. 

Unlike LMT methods using Bragg diffraction or Bloch oscillations, LMT Raman atom interferometers \cite{mcguirk2000large_LMTraman,HMuller_SDK_LMT_2018} need to reverse the direction, $\kappa$, of the effective wavevector between successive light pulses. The effective wavevector of both directions are present in \cite{HMuller_SDK_LMT_2018}, but the Doppler shift resolves the degeneracy of the Raman resonances for $\kappa=\pm1$. It is necessary therefore to switch rapidly between different frequencies during one interferometer sequence, while preserving phase coherence  \cite{Lee2022_gmotInterferometer}. In vertical accelerometers, the velocity of atoms falling under gravity naturally lifts the degeneracy. In horizontal accelerometers the atoms can be launched with enough velocity to lift the degeneracy \cite{xiaxi_thesis}, at the cost of some technical complexity. This complexity can be avoided by removing one of the retro-reflected Raman beams \cite{Sabulsky2019_chameleon, pengsensitive} but then it is not possible to reverse $\kappa$. \\ 

We present a novel method for achieving LMT in a Raman atom interferometer without the need for $k$-reversal. Our method closely follows the work in \cite{HMuller_SDK_LMT_2018}, but we eliminate the need for reversing $\kappa$ by employing a sequence of alternating microwave and optical transitions. This approach also removes the need to switch the frequency of the light pulses rapidly within a single interferometer sequence. It shares the benefit of being insensitive to ac Stark shifts as only $\pi$ pulses are used \cite{Weiss1994}. We demonstrate an implementation of this method and propose a pathway toward achieving larger momentum transfers, enabling even greater sensitivities.\\

\begin{figure*}
    \centering
    \includegraphics[width=0.9\linewidth]{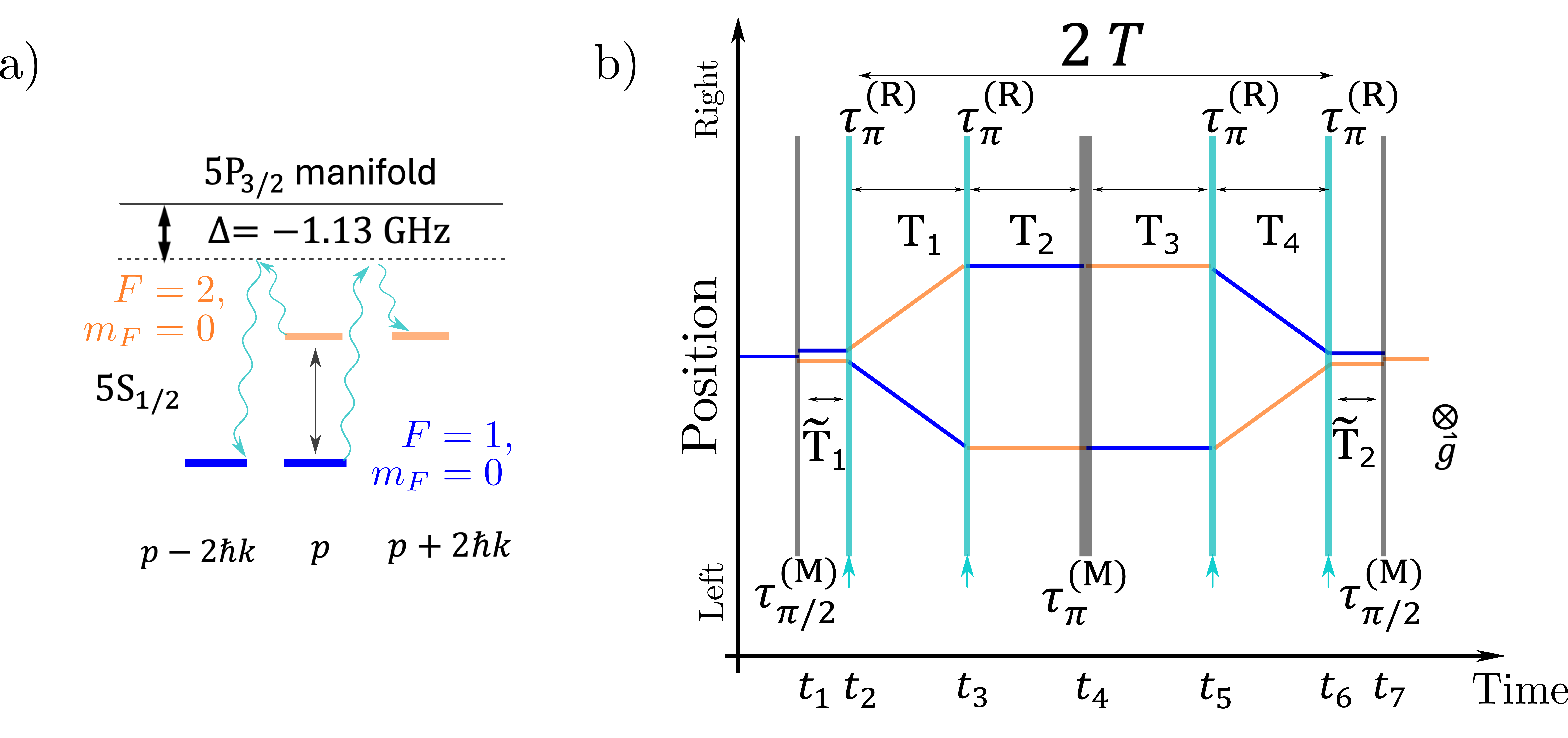}
    \caption{A 4$\hbar k$ large momentum transfer without $k$-reversal. (a) Energy-momentum states involved in the interferometer. We drive Raman transitions (cyan arrows) and microwave transitions (grey arrow) between $\ket{F=1,m_F=0}$ and $\ket{F=2,m_F=0}$ to induce spin-dependent kicks. (b)  A $4\hbar k $ large momentum transfer interferometer. The microwave interaction (grey lines) couples spin states of the same momentum state. The Raman interaction (cyan lines) couples different momentum states by spin-dependent momentum transfer.}
    \label{fig:principleDemonstration}
\end{figure*}

Our realization of the interferometer scheme uses $^{87}$Rb atoms. As shown in Fig. \ref{fig:principleDemonstration}a), we use microwave pulses to directly couple the hyperfine levels of the $(5s)^2 S_{1/2}$ state, and laser pulses detuned from the $(5p)^2P_{3/2}$ state to drive Raman transitions between these levels. Our 4$\hbar k$ LMT atom interferometer is realized by three microwave pulses and four Raman pulses, as is illustrated in Fig. \ref{fig:principleDemonstration}b). The interferometer sequence starts with a spin-dependent-kick beam splitter \cite{HMuller_SDK_LMT_2018}. This uses a microwave $\pi/2$ pulse followed by a  Raman $\pi$ pulse to create a momentum separation between the two interferometer arms of $4\hbar k$, where $k=2\pi/\lambda$ and $\lambda\approx780$ nm. More specifically, the microwave $\pi/2$ pulse puts each atom into an equal superposition of two spin states, $\ket{F,m_F} = \ket{1,0}$ and $ \ket{2,0}$, without changing its momentum state.  After a period  $\widetilde{T}_1$ of free evolution, a Raman $\pi$ pulse imparts momenta of $+2\hbar k$ to $\ket{1,0}$ and  $-2\hbar k$ to $\ket{2,0}$, as well as reversing the spin states. A second Raman $\pi$ pulse is applied after a further time $T_1$. This keeps the spatial separation between  two interferometer arms at a fixed distance, $d=4\hbar k T_1/M $, where  $M$ is the atomic mass. The atoms then evolve freely for a time $T_2$ before the spin states of the two arms are swapped by a microwave $\pi$ pulse. Subsequently, there is another free evolution time $T_3$, following which the two interferometer arms are closed, \textit{without $k$-reversal}, by applying two more Raman $\pi$ pulses separated by $T_4=T_1$. Finally, after freely evolving for $\widetilde{T}_2$, the two arms are made to interfere by applying the last microwave $\pi/2$ pulse.\\

After the interferometer sequence described above, the population in $\ket{2,0}$, $P_{F=2}$, is
\begin{equation}
    P_{F=2} = B-A\cos{\left(\Delta\phi^{\textrm{M}}+\Delta\phi^{\textrm{R}}\right)},
    \label{eq:p2}
\end{equation}
where $B$ is the middle point of the interference fringe, $A$ is the amplitude of the fringe and $\Delta\phi^{\textrm{M}}$ and $\Delta\phi^{\textrm{R}}$ are the phase differences along the two paths due to interactions with the microwave and Raman fields respectively.\\

The full derivation of $\Delta\phi^{\textrm{M}}$ and $\Delta\phi^{\textrm{R}}$ can be found in our supplemental material \cite{supplementaryMaterial}. Here we choose $T_1 = T_4$ in order to close the interferometer. The microwave phase difference is then,
\begin{equation}
    \Delta\phi^{\textrm{M} } = \left(\omega^{\textrm{M} }-\omega_0\right) (\widetilde{T}_1+T_2-T_3-\widetilde{T}_2),
\end{equation}
where $\omega^{\textrm{M}}$ is the microwave frequency and $\omega_0$ is the hyperfine splitting of the $(5s)^2S_{1/2}$ state. We choose a symmetric timing of $\widetilde{T}_1=\widetilde{T}_2$ and $T_2=T_3$ to ensure that $\Delta\phi^\mathrm{M}$ is strongly suppressed. With this timing, the Raman phase, calculated in the short-pulse limit  \cite{Kasevich1992_atomicFountain,hogan2009light}, is  
\begin{equation}
    \Delta\phi^{\textrm{R}}= a\hspace{0.25 em} \frac{T_1}{T} \left(2-\frac{T_1}{T}\right) 4 k \hspace{0.1 em}  T^2 =a \hspace{0.1 em} n \hspace{0.1 em} k \hspace{0.1 em} T^2,
    \label{eq:lmt_phase}
\end{equation}
where $a$ is the constant acceleration of the atom relative to the retro-reflection mirror, projected onto the axis of Raman beam propagation. The $4 k$ factor originates from the $4\hbar k$ differential momentum transfer per Raman pulse, and $2T=T_1+T_2+T_3+T_4$ is the interval between the first and the last Raman pulse. The dimensionless factor $n$ is a simple way of characterizing the sensitivity to acceleration for a given value of  $T_1/T$. We  note that, in the limit of $T_1/T\rightarrow 0$, Eq.(\ref{eq:lmt_phase}) gives $n_{\textrm{LMT}}\rightarrow 0$ and the interferometer has no acceleration sensitivity. In this limit, the two paths enclose no area and the interferometer becomes a microwave spin-echo sequence. In the limit of $T_1/T\rightarrow 1$, the interferometer reaches its maximum acceleration sensitivity of $n_{\textrm{LMT}} \rightarrow 4$ as the space-time area enclosed is the maximum available with a $4\hbar k$ momentum transfer.
In this Letter, we study the phase, $\Delta\phi^{\textrm{R}}$, defined in Eq. \ref{eq:lmt_phase}, and compare it with that of a standard three-pulse Mach-Zehnder atom interferometer.\\

The apparatus is very similar to that described in Sabulsky \textit{et al. }\cite{Sabulsky2019_chameleon}. Here we give a brief description of the experimental sequence. Approximately $10^8$ atoms of $^{87}$Rb are cooled to $\sim$$10 \,\mathrm{\upmu K}$ and optically pumped into the $(5s)^2 S_{1/2}$ $F = 1$ ground state, distributed across the three $m_F$ sublevels. A state preparation sequence, outlined in the supplemental materials and in \cite{pengsensitive}, drives approximately $5 \times 10^6$ of the atoms  into the magnetically insensitive $\ket{1,0}$ state and removes the rest. During the interferometer sequence, we apply a bias magnetic field to lift the degeneracy of the $m_F$ states. Laser light is provided by the $\mu$Quans system described in \cite{Sabulsky2019_chameleon}. The Raman frequencies, $\omega_1$ and $\omega_2$, differ by $2 \pi \times $6.834 GHz in order to drive the  hyperfine transition $\ket{F, m_F} = \ket{1,0} \rightarrow \ket{2,0}$.  \\

\begin{figure}
    \centering
    \includegraphics[width=1.0\linewidth]{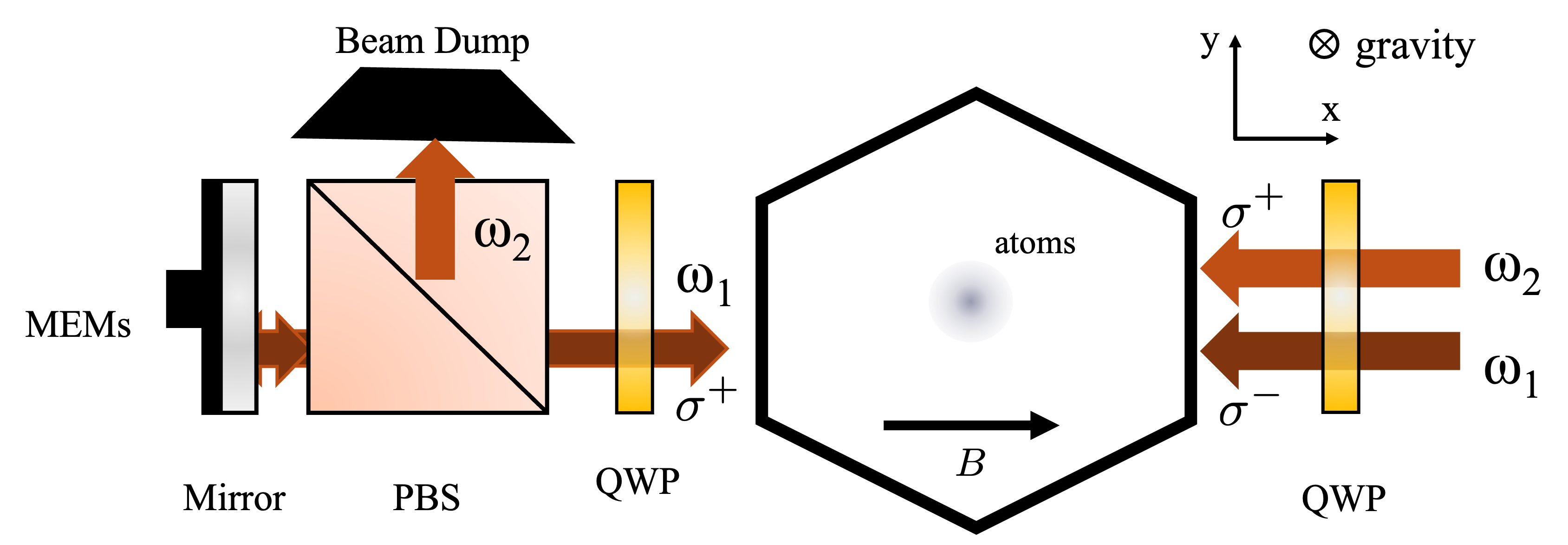}
    \captionsetup{justification=justified}
    \caption{Schematic of the Raman beam setup, showing the polarization scheme which rejects the counter-propagating $\omega_2$ light, to only drive the $\sigma^+-\sigma^+$ Raman transition. See main text for details. }
    \label{fig:raman-schematic}
\end{figure}

Figure~\ref{fig:raman-schematic} shows the preparation of the light for the interferometer. The light at the two Raman frequencies $\omega_1$($\omega_2$) leaves the output fiber with orthogonal linear polarizations, before passing through a quarter-wave plate (QWP) to become right-handed (left-handed) circular polarizations that drive $\sigma^-$($\sigma^+$) transitions. After passing through another QWP, $\omega_2$ is directed to a beam dump by a polarizing beam splitter (PBS) while $\omega_1$ passes through, reflects from the mirror, passes back through the PBS and the QWP resulting in a right-handed circular polarization, which completes the $\sigma^+-\sigma^+$ Raman transition. As $\omega_2$ is rejected, the $\sigma^--\sigma^-$ Raman transition is not driven \cite{RamanTransitionsFootnote}. A horn delivers microwave radiation to the atoms through a vacuum window. The microwave source is derived from the $\mu$Quans Raman laser system so that the radiation is phase-coherent with the beat note between the two Raman beams, as is necessary to form the interferometer. A micro electro-mechanical (MEMS) accelerometer is attached to the rear of the mirror to monitor its vibrations.\\

The interferometer scheme is shown in Fig.~\ref{fig:principleDemonstration}b), with  
 $\widetilde{T}_1=\widetilde{T}_2=1$ ms and $T_1 \cdots T_4$, all set to  4 ms. The fraction of atoms in the $F=2$ state, $P_{F=2}$, is determined by state-sensitive fluorescence detection.\\

\begin{figure}
\centering
\includegraphics[width=0.99\linewidth]{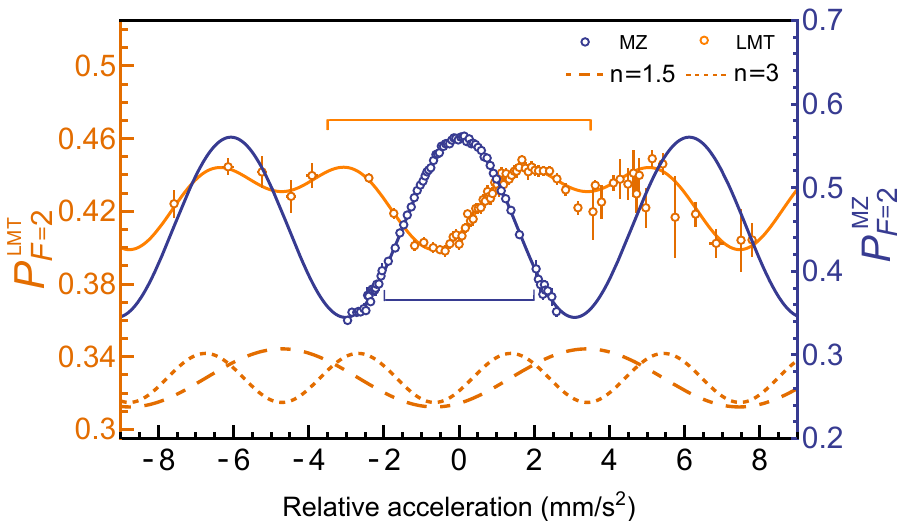}     
\caption{Atom interferometer fringes with $T=8\,$ms, shifted to be centred near zero acceleration. Orange data points: LMT interferometer. Blue points: three-pulse Mach-Zehnder interferometer. In the range indicated by the orange (blue) bracket, each orange (blue) point is the average of 40 measurements. Outside those ranges each point is the average of four measurements. Solid curves: fits to Eq.(\ref{eq:PMZ}) and Eq.(\ref{eq:PLMT}). Dashed and dotted curves: separate contributions in the fit from main and parasitic interferometers respectively (arbitrarily lowered for clarity). 
}
\label{fig:LMTvsMZ}   
\end{figure}

The orange points in Fig.\ \ref{fig:LMTvsMZ}, show $P_{F=2}$ measured as a function of the mirror acceleration. Each point in the plot is an average of either 40 or 4 measurements, and the error bar in $P_{F=2}$ represents the standard deviation of the mean of those measurements. No error bars are shown for the acceleration as these are all less than \qty{0.5}{mm/s^2}. For each measurement, the mirror acceleration $\langle a(t)\rangle_{\textrm{LMT}}$ is averaged over the time between the first and last Raman pulses, with a trapezoidal weighting,  as described in the supplemental material \cite{supplementaryMaterial} (see also references \cite{storey1994feynman,hogan2009light,barrett2014mobile} therein).  The range of accelerations plotted in Fig.\ \ref{fig:LMTvsMZ} is passively sampled by the free movements of the laser table and the mirror. In order to achieve a scan this wide, the table was not floated.

For comparison, the blue points in Fig.\ \ref{fig:LMTvsMZ} are taken using a three-pulse Mach-Zehnder (MZ) sequence with the same $T=8\,$ms and the standard triangular weighting function to obtain the time-averaged mirror acceleration $\langle a(t)\rangle_{\textrm{MZ}}$ \cite{barrett2014mobile}. These data were taken with the table floating.
We fit the MZ interference fringe to the function 
\begin{equation}
P^{\textrm{MZ}}_{F=2} = B-A\cos{\left(n_{\textrm{MZ}}\, \langle a(t)\rangle_{\textrm{MZ}}\, k\, T^2+\phi_0\right)},
 \label{eq:PMZ}
\end{equation}
where $\phi_0$ is an offset phase, with the result plotted in Fig.~\ref{fig:LMTvsMZ}. This gives $n_{\textrm{MZ}}=2.00\pm 0.01$, which is in excellent agreement with the expected value of $2$ due to the $2\hbar k$ momentum transfer per Raman pulse. \\

\begin{figure}
    \centering
    \includegraphics[width=0.9\linewidth]{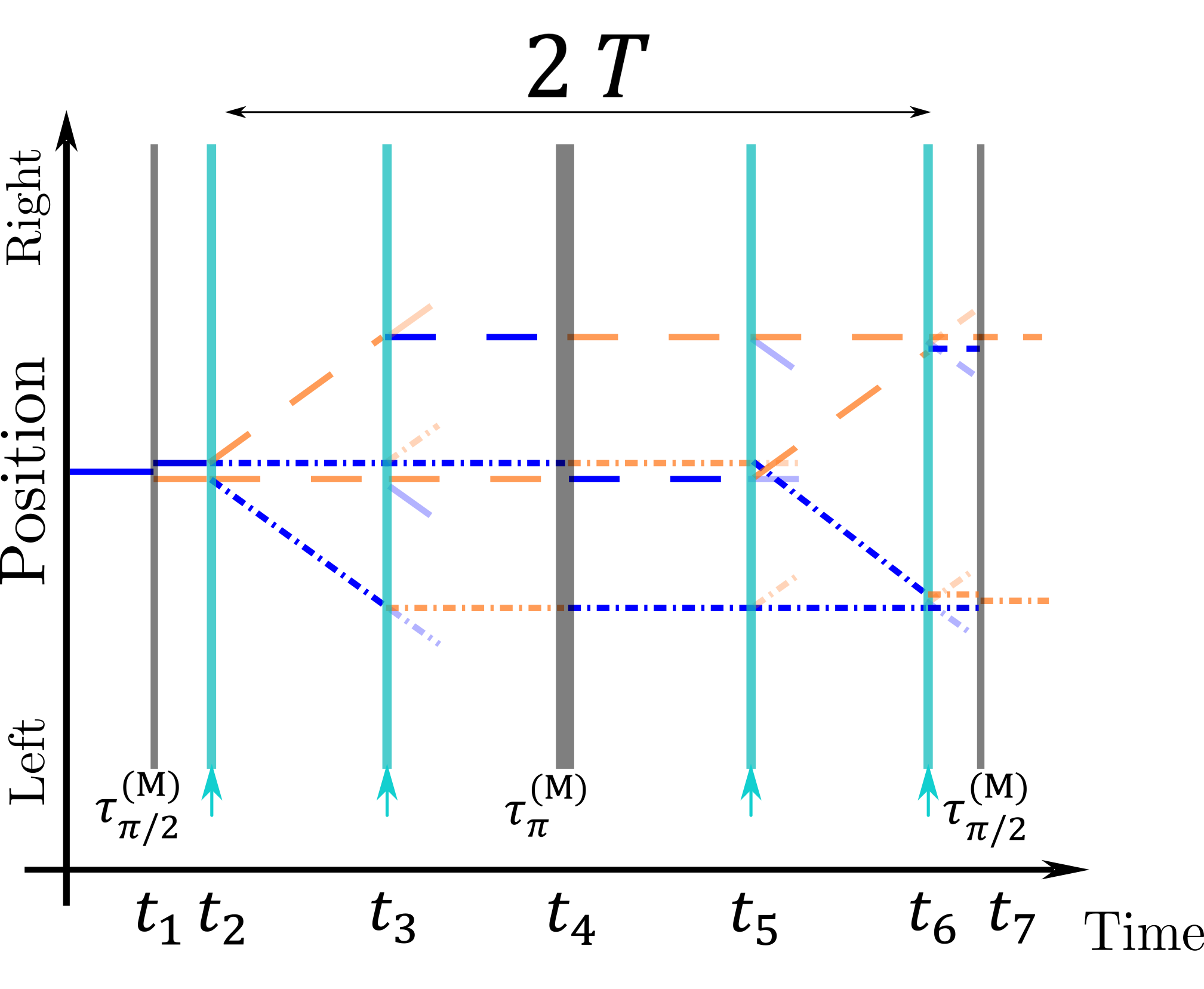}
    \caption{Parasitic interferometers in the scheme of Fig.\ref{fig:principleDemonstration}b). These form because the Raman pulse is not a perfect $\pi$ pulse. The dotted/dashed lines show the two closed paths with zero-momentum output states. Orange (blue) lines indicate $F=2$ ($F=1$) states. Faint lines indicate the starts of other paths.}
    \label{fig:parasites}
\end{figure}

In contrast, it is clear from the orange data in Fig.~\ref{fig:LMTvsMZ} that the LMT interferometer does not exhibit the single sinusoidal behavior proposed in Eq.(\ref{eq:p2}). This more complicated fringe pattern is due to the two additional parasitic interferometers shown in Fig.~\ref{fig:parasites}, which arise because the Raman pulses are not exactly $\pi$ pulses. This is unavoidable as the intensity of the Raman beams varies across the cloud of atoms. These parasitic interferometers require the same trapezoidal acceleration weighting function as the LMT interferometer, but have only half the acceleration sensitivity because they enclose half the area. More parasitic interferometers are created in a similar way by the other imperfect   Raman pulses, but they involve higher momenta that are outside the linewidth of the Raman transition, and are too weak to see. We note that the two arms of the main $4\hbar k$ LMT interferometer have opposite two-photon recoil shifts of $\pm 2\pi\times15$ kHz. Our Raman coupling of $2\pi\times50$ kHz is just sufficient to  drive transitions in both of these arms simultaneously.\\

In order to describe this more complicated fringe, we fitted the data to the function
\begin{equation}
\begin{split}
P^{\textrm{LMT}}_{F=2} =   B &-A_1\cos{\left(n^{(1)}_{\textrm{LMT}}\, \langle a(t)\rangle_{\textrm{LMT}} k\, T^2  +\phi_1\right)} \\
& -A_2\cos{\left(n^{(2)}_{\textrm{LMT}}\, \langle a(t)\rangle_{\textrm{LMT}} k\, T^2+\phi_2\right)},
\end{split}
 \label{eq:PLMT}
\end{equation}
where $A_i$, $B$, $n^{(i)}_{\textrm{LMT}}$, and $\phi_i$ are fitting parameters ($i$=1 for the desired interferometer and $i=2$ for the parasitic ones).
The fit, plotted in Fig.~\ref{fig:LMTvsMZ}, gives $n^{(1)}_{\textrm{LMT}}=3.00 \pm 0.05$ in good agreement with the value expected for the main interferometer by putting $T_1/T=\tfrac{1}{2}$ into Eq.(\ref{eq:lmt_phase}). This result confirms that the LMT interferometer does indeed have a finer fringe spacing than that of the three-pulse Mach-Zehnder interferometer by a factor of $1.5$  for the same time $T$. It also verifies that the trapezoidal acceleration weighting function is valid. We expect $n^{(2)}_{\textrm{LMT}}=\tfrac{1}{2}n^{(1)}_{\textrm{LMT}}$ because the parasitic interferometers enclose half the area, and that is confirmed by the same fit, which yields $n^{(2)}_{\textrm{LMT}}=1.50 \pm 0.04$.\\ 

 The dashed and dash-dotted lines in Fig.~\ref{fig:LMTvsMZ} show the separate contributions to the fit from the main and parasitic interferometers. We see that the minima of the parasitic fringes coincide with every second minimum of the main interferometer. This is to be expected if $\phi_1$ and $\phi_2$ simply represent our arbitrary choice of zero relative acceleration. In that case we would expect $\phi_1=2\phi_2$, and indeed the fit gives $1.01\pm0.06$ rad and $0.51\pm0.06$ rad. This proportionality serves as a useful test to show that our interferometer does not suffer from phase errors that depend differently on the enclosed area, e.g. light shifts or Zeeman shifts.\\

\begin{figure*}
    \centering
    \includegraphics[width=0.9\textwidth]{figures/fig4_subfigs_labelled_v4.png}
    
    \caption{Proposed extension from Fig.~\ref{fig:principleDemonstration}b) to a 4$N\hbar k$ interferometer without $k$-reversal. a) Schematic of the pulse timing. The two interferometer arms are opened and closed by microwave $\pi/2$ pulses (thin grey lines). All the Raman pulses (cyan solid lines) are $\pi$ pulses to induce spin-dependent momentum transfer, before microwave $\pi$ pulses (thick grey lines) swap their internal states without affecting the momentum. Note the paired Raman pulses at the center of the second and fourth sections (highlighted in red). These reverse the order of the alternating Raman and microwave sequence, as is necessary to reverse the acceleration between the two arms. The central microwave $\pi$ pulse not only equalises the time spent by each spin state on two arms, but also closes the interferometer without $k$-reversal. b) Energy-momentum states of one of the interferometer arms during part of the pulse sequence. The two-step transfer is analogous to a Bragg transition.}
    \label{fig:4Nhbark}
\end{figure*}

It is  instructive to compare the peak-to-peak amplitudes of the Mach-Zehnder and LMT fringes in Fig.\ \ref{fig:LMTvsMZ}. The former uses three Raman pulses: $\pi/2$, $\pi$, $\pi/2$. Averaged over the cloud of atoms, the Raman $\pi$ pulse makes a transition probability of only $0.55$. As a rough estimate, we can therefore expect a peak-to-peak amplitude of order $(0.55)^3$, which is close to the 0.2 that we observe. In contrast, the LMT interferometer requires seven pulses: three microwave pulses, for which the population transfer efficiency is $\sim\,$0.8, and four Raman $\pi$ pulses. The rough estimate of $(0.8)^3\times(0.55)^4$ is similar to the measured peak-to-peak amplitude of $\sim\,$0.03. \\

Figure~\ref{fig:4Nhbark} shows how to extend our method to larger recoils of $N\times 4\hbar k$ without $k$-reversal. Part (a) shows the overall pulse scheme, while (b) shows how the paired Raman and microwave pulses increase the momentum. The analogy with Bragg transitions is discussed in the supplemental material. The numbers of microwave and Raman pulses scale as $(4N-1)$ and $4N$ respectively, so both will need to be more efficient. At present, the microwave field is inhomogeneous because of standing waves formed by reflections within the  vacuum chamber, but with a  redesign, one could approach the ideal of a plane wave perpendicular to the Raman axis.  The  Raman $\pi$ pulse efficiency is low because of the intensity variations across the Gaussian laser beam. This difficulty can be removed by using the adiabatic-passage technique described in \cite{HMuller_SDK_LMT_2018}, which overcomes both the intensity variations and the opposite recoil shifts of the two interferometer arms. With a pulse efficiency of $96\%$, similar to that shown in \cite{HMuller_SDK_LMT_2018}, one can expect our scheme to achieve a recoil of $16\hbar k$ and a fringe peak-to-peak amplitude of $\sim 0.3$, which is similar to the amplitude in \cite{HMuller_SDK_LMT_2018}.\\

The resulting improvement in sensitivity to accelerations  would be valuable in several applications. Examples include compact low-drift accelerometers for navigation \cite{KasevichGroup2013_InertialLaunching, d’ArmagnacdeCastanet2024_rotationMeasurements, geiger2011_InertialNavigation, cheiney2018_InertialNavigation, Templier2022_InertialSensing} and fundamental physics experiments, such as gravitational wave detection using a pair of horizontal accelerometers \cite{Zhan2020_ZAIGA, Canuel2018_MIGA}. In these applications, our method allows enhancing acceleration sensitivity without rapidly switching $\kappa$ in one measurement, however, it will be useful to reverse $\kappa$ from one measurement to the next. The two phases $\left(+ n k a T^2+\varphi\right)$ and $\left(- n k a T^2+\varphi\right)$ would reveal the presence of any unwanted phase $\varphi$ that does not depend on the sign of $\kappa$.\\

In summary, we have demonstrated an approach that removes the need to reverse $\kappa$ when applying spin-dependent momentum kicks in atom interferometry \cite{HMuller_SDK_LMT_2018}. Our approach still retains the benefits of the original proposal, namely (i) insensitivity to the ac Stark shifts by limiting the light to $\pi$-pulses and (ii) encoding of the momentum states in the internal states to enable readout.\\

\begin{acknowledgments}
We acknowledge engineering support by J. Dyne, software support by Teodor B Krastev, and inspiring discussions with Malo Cadoret (Conservatoire National des Arts et Métiers, Paris France).   We thank the UK Science and Technology Facilities Council (STFC) for funding through grants ST/W006316/1 and ST/V00428X/1.

\end{acknowledgments}

\bibliography{references}

\begin{thebibliography}{51}
\bibitem [{Ram()}]{RamanTransitionsFootnote}%
  \BibitemOpen
  \href@noop {} {}\bibinfo {note} {When the Raman beams are horizontal, the $\sigma^+$-$\sigma^+$ and $\sigma^-$-$\sigma^-$ Raman transitions are degenerate in frequency and have opposite recoil momenta. It is therefore necessary to suppress one of them to avoid splitting the wavefunction at each $\pi$-pulse. It is not necessary if the beams are vertical because then the Doppler shift removes the frequency degeneracy.}\BibitemShut {Stop}%
\end{thebibliography}


\clearpage

\section{Supplemental material for: A large-momentum-transfer Raman atom interferometer without $k$-reversal}

In an atom interferometer, the probability of detecting atoms in one of two output states is determined by the difference in the phases, $\Delta\phi$, accumulated along the two arms. Our $\Delta\phi$ is proportional to the acceleration of the atoms relative to the mirror that retro-reflects the laser light. Vibrations of the mirror are fast enough that the acceleration changes during the interferometer sequence, and therefore we need to consider what kind of average determines $\Delta\phi$. In the case of a standard three-pulse Mach-Zehnder (MZ) interferometer, the acceleration is averaged with a triangular weighting factor \cite{barrett2014mobile}. Here we derive the appropriate weighting function for our $4\hbar k$ large momentum transfer (LMT) interferometer and consider how to generalise it to other schemes.\\

To derive the weighting, we firstly review each of the three phase contributions to the total phase difference, $\Delta\phi_{\textrm{LMT}}$, for our $4\hbar k$ LMT interferometer. 
We then show the acceleration weighting function of our $4\hbar k$ LMT interferometer is trapezoidal rather than a triangle. Finally, we provide a simple picture to understand the shape of an acceleration weighting function. This picture provides an intuitive way to obtain the acceleration weighting function for an arbitrary pulse sequence, such as a $4N\hbar k$ LMT interferometer proposed in Fig. 5 in the main text.

\section{Phase contributions}
Assuming all the $^{87}\textrm{Rb}$ atoms are prepared in the F=1 state before the interferometer pulses, then the final population in F=2, $P_{F=2}$, at the output of our $4\hbar k$ interferometer is given by 
\begin{equation}
    P_{F=2} = B-A\cos{\left(\Delta\phi\right)},
    \label{eq:populationInterferometerPhase}
\end{equation}
where $B$ is the middle point of the interference fringe, $A$ is the amplitude of the fringe and  $\Delta\phi$ is the total phase difference between two interferometer arms. 

Following the formalism and notation in \cite{hogan2009light}, we can calculate this total phase difference, $\Delta\phi$, by separating it into three contributions
\begin{equation}  \Delta\phi=\Delta\phi^{\textrm{prop}}+\Delta\phi^{\textrm{sep}}+\Delta\phi^{\textrm{int}},
    \label{eq:interferometerPhase}
\end{equation}
where $\Delta\phi^{\textrm{prop}}$ is the difference of the propagation phase between the two interferometer arms, $\phi_{\ell}^{\textrm{prop}}$ and $\phi_{\textrm{r}}^{\textrm{prop}}$.  We use the same ${\ell}/{\textrm{r}}$ (left/right) notation for the separation phase difference, $\Delta\phi^{\textrm{sep}}$,  and the interaction phase difference, $ \Delta\phi^{\textrm{int}}$.\\

We note that $\Delta\phi^{\textrm{sep}}=0$ for our interferometer, as we have chosen $T_1=T_4$ such that for the final pulse the two interferometer arms overlap in space, closing it. Thus, we only need to calculate the sum of $\Delta\phi^{\textrm{prop}}$ and $\Delta\phi^{\textrm{int}}$ to evaluate $\Delta\phi$.

\subsection{Propagation phase}
The propagation phase originates from the free-evolution of the  wave packet and it is given by  
\begin{equation}  
    \Delta\phi^{\textrm{prop}}=\underbrace{\sum_{\textrm{right}} \int_{t_I}^{t_F} \frac{L_c-E_i}{\hbar}  dt}_{\phi^{\textrm{prop}}_{\textrm{r}}}-\underbrace{\sum_{\textrm{left}} \int_{t_I}^{t_F} \frac{L_c-E_i}{\hbar} dt}_{\phi^{\textrm{prop}}_{\ell}},
    \label{eq:interferometerPhase_propagation}
\end{equation} 
where the sum is over all segments of the left (or right) path, $L_c$ is the classical Lagrangian evaluated at the centre of mass of the wavepacket on the left (or right) trajectory, $E_i$ is the internal atomic energy on that trajectory, and  $t_I$ and $t_F$ are the initial time and the final time of each path segment.\\

In our interferometer, the contribution of  $L_c$ is found to be zero. Therefore, $\Delta\phi^{\textrm{prop}}$ is determined by the energy difference of the two internal states, $\omega_0=(E_2-E_1)/\hbar$, and the  time spent in each internal state, given by
\begin{equation}  
    \Delta\phi^{\textrm{prop}}=-\omega_0 \,(\tilde{T}_1-T_1+T_2-T_3+T_4-\tilde{T}_2)
    \label{eq:interferometerPhase_propagationResult},
\end{equation}
where $\tilde{T}_1, T_1, T_2, T_3, T_4$ and $\tilde{T}_2$ are the temporal separations between interferometer pulses shown in Fig. 1b in the main text.

\subsection{Interaction phase}
The interaction phase comes from the phase of the microwave field or Raman laser imprinted on the wavepacket during a coherent population transfer. The difference in the interaction phases accumulated on the two  interferometer arms is given by

\begin{equation}
    \begin{split}
    \Delta\phi^{\textrm{int}} = &\underbrace{\sum_{j}\pm\phi^{\textrm{int}}\big(x_{\textrm{r}}(t_j),t_j\big)}_{\phi_{\textrm{r}}^{\textrm{int}}}-\underbrace{\sum_{j}\pm\phi^{\textrm{int}}\big(x_{\ell}(t_j),t_j\big)}_{\phi_{\ell}^{\textrm{int}}},
\end{split}
\label{Eq.interactionPhaseDiff}
\end{equation}
where the summation is over all the interaction events, which occur at times $t_j$ and at positions $x_{\textrm{r}}(t_j)$ and $x_{\ell}(t_j)$ on the classical trajectories. We will further split the interaction phase $\phi^{\textrm{int}}$ out according to whether it arises from a microwave transition or a Raman transition, $\phi^{\textrm{M}}$ and $\phi^{\textrm{R}}$ respectively. The interaction phase has a $+$($-$) sign when a photon is absorbed(emitted) by an atom.\\

The microwave phase at an atom's center of mass position $\textbf{x}_c$ at time $t$ is given by 
\begin{equation}
\phi^{\textrm{M}} = \omega^{\textrm{M}} t - \textbf{k}^{\textrm{M}}\cdot(\textbf{x}_c(t)-\textbf{x}_{\textrm{horn}})+\varphi^{\textrm{M}},
\label{Eq.mwPhase}
\end{equation}
where $\omega^{\textrm{M}}$,  $\textbf{k}^{\textrm{M}}$ and $\varphi^{\textrm{M}}$ are the frequency, wavevector, and a constant initial phase of the microwave field. We assume the position of the horn $\textbf{x}_\textrm{horn}$ is static. 

Similarly, the Raman phase is given by 
\begin{equation}
\phi^{\textrm{R}}=\omega^{\textrm{R}} t - \textbf{k}^{\textrm{R}}\cdot \big(\textbf{x}_c(t)-\textbf{x}_{\textrm{mirror}}(t)\big)+\varphi^{\textrm{R}},
\label{Eq.ramanPhase}
\end{equation}
where $\omega^{\textrm{R}}$, $\textbf{k}^{\textrm{R}}$ and $\varphi^{\textrm{R}}$ are the frequency, wavevector and a constant initial phase of the Raman light. $\omega^{\textrm{R}}=\omega_1-\omega_2$ is the frequency difference between the two Raman lasers and $\textbf{k}^{\textrm{R}}=\textbf{k}_{1}-\textbf{k}_{2}$ has a magnitude of $2k$, where $k=2\pi/\lambda$ is the averaged wavenumber of two Raman lasers of wavelength $\lambda\approx 780 \, \textrm{nm}$. The position of the retroreflection mirror $\textbf{x}_\textrm{mirror}$ can vary in time. Its acceleration is monitored by the MEMS accelerometer.\\

In our $4\hbar k$ interferometer, the interaction phases accumulated along each arm are given by 
\begin{equation}
\begin{split}
\phi^{\textrm{int}}_r = & \phi^{\textrm{R}}_{2,r}-\phi^{\textrm{R}}_{3,r}+\phi^{\textrm{M}}_{4,r}-\phi^{\textrm{R}}_{5,r}  +\phi^{\textrm{R}}_{6,r},\\
\phi^{\textrm{int}}_\ell = & \phi^{\textrm{M}}_{1,\ell}-\phi^{\textrm{R}}_{2,\ell}+\phi^{\textrm{R}}_{3,\ell}-\phi^{\textrm{M}}_{4,\ell}+\phi^{\textrm{R}}_{5,\ell}
 -\phi^{\textrm{R}}_{6,\ell}+\phi^{\textrm{M}}_{7,\ell},
\end{split}
\label{Eq.interectionPhaseLeftRight}
\end{equation}
where we have labeled each phase contribution $\phi^{\textrm{X}}_{j,\alpha}$ with X denoting Raman or MW, $j$ denoting pulse number, and $\alpha$ the left or right arm. Substituting equations Eq.\ref{Eq.mwPhase} to 
Eq.\ref{Eq.interectionPhaseLeftRight} into Eq.\ref{Eq.interactionPhaseDiff} gives the interaction phase difference between the two arms of our interferometer:
\begin{equation}
    \begin{split}
    \Delta\phi^{\textrm{int}} = & \omega^{\textrm{M}}(-t_1+2t_4-t_7) \\
    &
    +\textbf{k}^{\textrm{M}}
    \cdot
    \left(
    \textbf{x}_{1,\ell}^\textrm{M}
   -\textbf{x}^\textrm{M}_{4,\ell} -
    \textbf{x}^{\textrm{M}}_{\textrm{4,r}} +
    \textbf{x}^{\textrm{M}}_{\textrm{7,r}}
    \right)
    \\
    &+2\omega^{\textrm{R}}(t_2-t_3-t_5+t_6)
   \\    &-2\textbf{k}^{\textrm{R}}\cdot\big(\bar{\textbf{x}}_{\textrm{2}}^{\textrm{R}}-\bar{\textbf{x}}_{\textrm{3}}^{\textrm{R}}-\bar{\textbf{x}}_{\textrm{5}}^{\textrm{R}}+\bar{\textbf{x}}_{\textrm{6}}^{\textrm{R}}\big),
    \end{split}
\label{eq:interferometerPhase_interacionPhase}
\end{equation}
where we use the averaged classical atom position relative to the mirror, $\bar{\textbf{x}}_{\textrm{j}}^{\textrm{R}}$: 
\begin{equation}
\begin{split}
\textbf{x}^{\textrm{R}}_{\ell,r}(t_j)&=\textbf{x}_{\ell,r}(t_j)-\textbf{x}_{\textrm{mirror}}(t_j), 
\\[8pt] 
\bar{\textbf{x}}_{\textrm{j}}^{\textrm{R}}&=
\frac{\textbf{x}^{\textrm{R}}_\ell(t_j)+\textbf{x}^{\textrm{R}}_r(t_j)}{2}.
\label{Eq.relativePosition}
\end{split}
\end{equation}

\subsection{Total phase difference between two arms}

The total phase difference between the two interferometer arms can be calculated by substituting Eq.\ref{eq:interferometerPhase_propagationResult} and Eq.\ref{eq:interferometerPhase_interacionPhase} into Eq.\ref{eq:interferometerPhase}. Noting that we closed the interferometer by choosing $T_1=T_4$, we can separate the microwave and Raman contributions and express $\Delta\phi$ as 
\begin{equation}
    \Delta\phi = \Delta\phi^{\textrm{M}}+\Delta\phi^{\textrm{R}},
    \label{eq.Phase_MW_Raman}
\end{equation}
which is the same form as that of  the Eq. 1 in the main text. Then, the microwave phase difference and the Raman phase difference are
\begin{equation}
    \begin{split}
        \Delta\phi^{\textrm{M}}=& \, (\omega^{\textrm{M}}-\omega_0)\left((\tilde{T}_1+T_1+T_2-(T_3+T_4+\tilde{T}_2)\right) \\
    &+\textbf{k}^{\textrm{M}}\cdot\big(\textbf{x}_{1,\ell}^{\textrm{M}}-\textbf{x}^{\textrm{M}}_{4,\ell}-\textbf{x}^{\textrm{M}}_{\textrm{4,r}}+\textbf{x}^{\textrm{M}}_{\textrm{7,r}}\big),\\[8pt]
 \Delta\phi^{\textrm{R}}=& \, 2\,(\omega^{\textrm{R}}-\omega_0)(T_4-T_1)\\
 &-2\textbf{k}^{\textrm{R}}\cdot\big(\bar{\textbf{x}}_{\textrm{2}}^{\textrm{R}}-\bar{\textbf{x}}_{\textrm{3}}^{\textrm{R}}-\bar{\textbf{x}}_{\textrm{5}}^{\textrm{R}}+\bar{\textbf{x}}_{\textrm{6}}^{\textrm{R}}\big).    
    \end{split}
    \label{eq.mwAndRamanPhases}
\end{equation}

We assume that in the horizontal direction, the atoms have some constant acceleration $a$, and that the scale of the horizontal trajectories, which is about 800 $\mu$m, is much less than the microwave wavelength of 4.4 cm. In addition, the standing wave form of the microwave in our system allows us to ignore the spatially dependent part of the microwave phase in Eq.~\ref{eq.mwAndRamanPhases} to re-write the microwave phase difference between the two arms as: 
\begin{equation}
    \begin{split}
        \Delta\phi^{\textrm{M}}=& \,\delta^{\textrm{M}}\,(\tilde{T}_1+T_2-T_3-\tilde{T}_2), 
    \end{split}
\end{equation}
where $\delta^{\textrm{M}}=\omega^{\textrm{M}}-\omega_0$ is the microwave detuning, and we have used $T_1=T_4$ for our closed interferometer. This is the derivation of Eq. 2 in the main text.\\

To derive Eq. 3 we assume a symmetric pulse sequence such that $\tilde{T}_1=\tilde{T}_2$ and $T_2=T_3$ in addition to $T_1=T_4$ required by a closed interferometer, which simplifies the Raman phase difference to 
\begin{equation}
    \begin{split}
\Delta\phi^{\textrm{R}}=&a\hspace{0.25 em} T_1 (T_1+T_2+T_3)\\
 =&a\hspace{0.25 em} \frac{T_1}{T} \left(2-\frac{T_1}{T}\right) 4 k \hspace{0.25 em}  T^2.   
    \end{split}
\end{equation}
We have used $T_2=T-T_1$ to show how $\Delta\phi$ depends on the separation between the first two Raman pulses, $T_1$, and the time between the first and the last Raman pulses, $2T$. Dividing $T_1$ by $T$ makes it simple to compare the acceleration sensitivities of the LMT and MZ interferometers.

\section{Acceleration weighting function}

Phase noise from vibrations has been well-studied \cite{barrett2014mobile} in a three-pulse MZ interferometer, where only the laser phase contributes to the total interferometer phase $\Delta\phi_{\textrm{MZ}}$. We apply the same formalism to our LMT interferometer. Although our interaction phase consists of both the Raman laser phase and the microwave phase, only the Raman laser phase is sensitive to the change of the Raman mirror position as shown in Eq.\ref{eq.mwAndRamanPhases}, and so we will only be considering those pulses when constructing our weighting function. To simplify the discussion we will assume 1D motion of atoms along the direction of the Raman beams.\\

If we consider an infinitesimal change in the Raman mirror position at time $t'$, $dx (t')$, we can express the relative positions of the atom, $\bar{x}_{j}^{\textrm{R}}$ defined in Eq. \ref{Eq.relativePosition}, to be
\begin{equation}
    \begin{split}
    &\bar{x}_{j}^{\textrm{R}}\rightarrow\bar{x}^{\textrm{R}}_{j}, \qquad \qquad \, \,\quad t_j<t'.\\
    &\bar{x}_{j}^{\textrm{R}}\rightarrow\bar{x}^{\textrm{R}}_{j}+dx (t'), \qquad t_j>=t'.
    \end{split}
\label{eq.changeInPosition}
\end{equation}
These changes lead to a change in the total interferometer phase, $d\Delta\phi$, which can be written using a time-domain position sensitivity function $g_{x}(t')$ defined by 
\begin{equation}
d \Delta\phi = g_{x}(t') dx(t'). 
\end{equation}
Substituting Eq.\ref{eq.changeInPosition} into Eq.\ref{eq.Phase_MW_Raman} yields 
  \begin{equation}
    g_x(t')=
    \begin{cases}
      +4k, & \text{if}\ t_2<t'<t_3 \\
      -4k, & \text{if}\ t_5<t'<t_6 \\
      0, & \text{otherwise}.
    \end{cases}
  \end{equation}

The vibration-induced phase shift can be calculated 
\begin{equation}
\begin{split}
    \delta(\Delta\phi) & = \int_{-\infty}^{+\infty}g_{x}(t') dx(t') \\
    & = \int_{-\infty}^{+\infty}g_{x}(t') \frac{dx(t')}{dt'} dt' = \int_{-\infty}^{+\infty}g_{x}(t') v(t') dt'.
\end{split}
\end{equation}
The position sensitivity function can also be regarded as a velocity weighting function.

If we define $g_{x}(t')$ as $dG(t')/dt'$ with a constraint that $G(\pm\infty)=0$, then $\delta(\Delta\phi)$ can be written in terms of a time-dependent acceleration by
\begin{equation}
\begin{split}
\delta(\Delta\phi) & = \int_{-\infty}^{+\infty}dG(t')\, v(t') \\
& = G(t') v(t')\bigg|_{-\infty}^{+\infty} - \int_{-\infty}^{+\infty}G(t')\, \frac{dv(t')}{dt'} dt' \\
    & = -\int_{-\infty}^{+\infty}G(t')\, a(t')\, dt'  \\
    & = 4 k \int_{-\infty}^{+\infty} f(t')\, a(t') dt',
\end{split}
\label{eq.accelerationWeight}
\end{equation}
where we have used integration by parts and taken $G(\pm\infty)=0$. This gives us an acceleration weighting function $f(t')=-G(t')/(4k)$, which, with the substitutions $t_3-t_2=T_1$ and $t_6-t_5=t_3-t_2$, is given by
\begin{equation}
    f(t')=
    \begin{cases}
      -t', & \text{if}\ t_2<t'<t_3 \\
      -T_1, & \text{if}\ t_3<t'<t_5 \\
      (t'-t_5)-T_1, & \text{if}\ t_5<t’<t_6 \\
      0, & \text{otherwise}.
    \end{cases}
\end{equation}
The weighting function $f(t')$ has a trapezoid shape for our LMT interferometer, which is different from a triangular acceleration weighting function for a standard MZ interferometer.

We note that $f(t')$ is not normalized. We define a normalization integration $\mathcal{N}=\int_{-\infty}^{+\infty} f(t')dt'$ to calculate our weighted average acceleration

\begin{equation}
    \langle a \rangle =  \int_{-\infty}^{+\infty} \frac{f(t')}{\mathcal{N}}\, a(t') dt'.
\end{equation}
where $f(t')/\mathcal{N}$ is a normalized weighting function of a trapezoid shape. This is how we calculated the time-average acceleration, $\langle a(t) \rangle_\textrm{LMT}$, in the main text.

\section{Geometric explanation for the shape of acceleration weighting functions}
 
In this section, we provide a simpler explanation for the shape of the acceleration weighting function, $f(t)$. This allows us to  obtain $f(t)$ directly, rather than deriving it from the position sensitivity function.

The change in the total phase difference between the two arms, $\delta\Delta\phi$, only depends on the change of the relative position of atoms referenced to the mirror position. Let us therefore calculate the same $\delta\Delta\phi$, not in the inertial  lab frame, but in the non-inertial mirror frame, where the mirror is stationary and the atoms have the additional vibrating motion. In this frame the Lagrangian of an atom, $\mathcal{L}^{\textrm{mirror}}$, has an additional term relative to its lab-frame Lagrangian: 
\begin{equation}\mathcal{L}^{\textrm{mirror}}=\mathcal{L}^{\textrm{lab}}-\underbrace{m\,a(t) \ x}_{\mathcal{L}_{\varepsilon}},
\end{equation}
where $\mathcal{L}_{\varepsilon}$ can be understood as a linear potential $V(x)=m\,a(t)\,x$ due to a homogeneous fictitious force $-m\, a(t)$. The lab frame Lagrangian  is simply given by  $\mathcal{L}^{\textrm{lab}}= m\ \dot{x}^2/2$. 

As proved in the path-integral approach to atom interferometry \cite{storey1994feynman},  $\mathcal{L}_{\varepsilon}=-m\,a(t) \ x$ can be treated as a perturbation because its dependence on the spatial coordinates is less than quadratic. Hence the  phase shift is obtained by integrating  the perturbed  Lagrangian, $\mathcal{L}_{\varepsilon}$, along the unperturbed path, $\tilde{x}(t)$: 
\begin{equation}
    \begin{split}
    \delta\Delta\phi&=\Delta \phi^{\textrm{prop}}_{\mathcal{L}_{\varepsilon}}\\&=\sum_{\textrm{right}} \int_{t_I}^{t_F} \frac{\mathcal{L}_{\varepsilon}}{\hbar}  dt' - \sum_{\textrm{left}} \int_{t_I}^{t_F} \frac{\mathcal{L}_{\varepsilon}}{\hbar} dt'\\
    &=-\int_{-\infty}^{+\infty} \frac{1}{\hbar} m\,\big(\tilde{x}_{\textrm{r}}(t')-\tilde{x}_{\ell}(t')\big)\,a(t')\,dt'.
    \end{split}
\label{eq.accWeightingFunctionFromPathIntegral}
\end{equation}
where the unperturbed path, $\tilde{x}_{\textrm{r}}$($\tilde{x}_{\ell}$), of the left (right) interferometer arm is determined by the lab frame Lagrangian, $\mathcal{L}^{\textrm{lab}}$.  These unperturbed paths $\tilde{x}_{\ell\textrm{,r}}$ are  illustrated in the Fig. 1 (b) in the main text.

Comparing Eq.\ref{eq.accWeightingFunctionFromPathIntegral} to Eq. \ref{eq.accelerationWeight}, we find 
\begin{equation}
    f(t)=-\frac{\big(\tilde{x}_{\textrm{r}}(t')-\tilde{x}_{\ell}(t')\big)}{4\hbar\,k/m}.
\end{equation}
 Thus, we see $f(t)$ is proportional to the separation of the two interferometer arms, which has a trapezoidal shape for our $4\hbar$k LMT interferometer and a triangular shape for a Mach-Zehnder interferometer. The same conclusion is also applicable for a $4N\hbar$k LMT interferometer illustrated in Fig. 4 in the main text.
 
\section{In relation to Bragg transisions}
 
Our scheme can be undertood as a two-step Bragg transition. Figure 5(b) in the main text shows in detail how the Raman transitions give momentum to the atom. Upon absorbing a photon from one of the Raman beams, an atom of momentum $p$  is stimulated to emit a photon of different frequency into the counter-propagating beam, thereby driving the transition $\ket{F=1,p}\rightarrow\ket{F=2,p+2\hbar k}$. The microwave pulse then returns the atom to its original internal state. The net effect is the same as that of a Bragg grating, where a photon is transferred from one beam to the other, but the internal state does not change. There, however, the photon frequencies differ only by the small recoil shift ($3.77$ kHz for Rb), and for pulses shorter than the inverse of this frequency, there can be a loss of atoms to higher diffraction orders \cite{muller2008atom, Beguin2023}. This is not an issue with the Raman scheme.

\section{Methods}
    
We load $10^8$  atoms from a $\textrm{2D}^{+}$-MOT \cite{Dieckmann1998TheTM}, to a 3D magneto-optical trap (MOT) of cooling light $\sim 16$ MHz red detuned with the $(5s)^2 S_{1/2}$ $F=2 \rightarrow$ $(5p)^2P_{3/2}$ $F' = 3$ transition, in addition to repump light near resonant with the $(5s)^2 S_{1/2}$ $F=1 \rightarrow$ $(5p)^2P_{3/2}$ $F' = 2$ transition. Over 2 ms we ramp the frequency of our cooling laser to $\sim 150$ MHz red detuned, then, after a further 2 ms, we ramp the intensity down to zero, releasing the atoms at near 10 $\mu$K. We then turn off the repump light, optically pumping the atoms into the $F = 1$ ground state, distributed across the three $m_F$ sublevels. Throughout the sequence, we apply we apply a Raman bias field of 0.38 G to lift the degeneracy of the $m_F$ states.\\

With the intensities of the Raman beams tuned to zero the light shift, and our atoms distributed across the $m_F = -1, 0, 1$ sublevels, we apply a Raman $\pi$ pulse to drive the $\ket{1,0} \rightarrow \ket{2,0}$ transition, before applying repumping light to pump the remaining atoms from $\ket{1,\pm 1}$ to $F=2$. Another Raman $\pi$ pulse transfers the atoms from $\ket{2, 0} \rightarrow \ket{1, 0}$, leaving the atoms in $\ket{F,m_F} = \ket{2,m_F \neq 0}$. We then apply the cooling light to selectively `blow away' the atoms in the $F=2$ state, leaving typically $25 \% $ of the initial population, almost all in the $\ket{1,0}$ state. We do this two-step state selection as it does an element of velocity selection, too, improving our Raman Rabi flops. \\

The interferometer scheme is shown in the main text Fig.1 b). Our Raman microwave $\pi/2$ pulse of $51$ $\mu$s places the atoms into a superposition of the two hyperfine states $\ket{1,0}$ and $\ket{2,0}$. The atoms evolve for 1 ms before we apply our first Raman $\pi$ pulse of $11$ $\mu$s, separating the arms of the interferometer. After $T_1$ = 4 ms, the second Raman $\pi$ pulse of $11$ $\mu$s reverses the states of the two arms and sets them on parallel trajectories, then after $T_2$ = 4 ms a microwave $\pi$ pulse of $100$ $\mu$s reverses their spin states while maintaining their momentum. After $T_3$ = 4 ms a Raman $\pi$ pulse of $12$ $\mu$s delivers another spin-dependent kick which overlaps the arms after $T_4$ = 4 ms, when we apply a final Raman $\pi$ pulse of $12$ $\mu$s. After a final 1 ms free evolution time we close the interferometer with a microwave $\pi/2$ pulse of $45$ $\mu$s. The ratio of atoms in $F = 2$ state, $P_F = 2$, at the end of the interferometer is determined by fluorescence detection.\\

The MEMS accelerometer (Innalabs, AI-Q-2120) is a microfabricated quartz pendulous servo accelerometer~\cite{beitia2015quartz}. Based on the specification, we expect a noise floor of \qty{0.24} {mm/s^2} for each shot of measurement. The MEMS accelerometer is  superglued to the back of the retro-reflection mirror, which tracks the vibrations of the mirror along the Raman axis. Said vibrations will be imprinted upon the atoms, and so by tracking the mirror, we are able to track the vibrations experienced by the atoms.

\end{document}